\documentclass[12pt]{article}
\usepackage{pqft}
\usepackage{graphicx}

\begin{document}
\title{Kinetic Theory of the Quantum Field Systems With Unstable Vacuum}
\author{S.~A.~Smolyansky\inst{1},~V.~V.~Skokov\inst{2},~A.~V.~Prozorkevich\inst{1}}
\institutes{\inst{1}
Physics Department, Saratov State University, 410071,
Saratov, Russia
\inst{2}
Bogoliubov Laboratory of Theoretical Physics, Joint Institute
for Nuclear Research, 141980, Dubna, Russia }

\maketitle

\vspace{3mm}
\abstract{ The description of quantum field systems with meta-stable vacuum
is motivated by studies of many physical problems (the decay of disoriented
chiral condensate \cite{SSP:SGU:Keiser}, the resonant decay of CP-odd
meta-stable states \cite{SSP:SGU:Ahrensmeier,SSP:SGU:Blaschke},
self-consistent model of QGP pre-equilibrium evolution \cite{SSP:SGU:we},
the phase transition problem in the systems with broken symmetry
\cite{SSP:SGU:Baacke} etc). A non-perturbative approach based on the kinetic
description within the framework of the quasi-particle representation was
proposed here. We restrict ourselves to  scalar field
theory. As an example, the models with potentials of polynomial type are
considered.}

\authorrunning{S.A.Smolyansky, V.V.Skokov, A.V.Prozorkevich}

\titlerunning{Kinetic Theory of the Quantum Field Systems ...}

\section{Introduction}

Separating of the classical background field is a well known method in
solving QFT problems in non-pertrubative approach
\cite{SSP:SGU:Grib,SSP:SGU:Birrel,SSP:SGU:Kluger}. In these methods the
quantum fluctuations can be described by perturbation theory. We consider a
particular class of scalar field theories with the background field, which
is strong enough to generate quasiparticles and propose here the kinetic
approach for the description of this process. The back reaction mechanism,
i.e. the particle production influence on background field is also
discussed. Using the oscillator representation, we derive the generalized
kinetic equation with non-pertrubative source term for description of
particle-antiparticle creation under action of background field and equation
of motion for it. The efficiency of this method was demonstrate previously
in the framework of scalar QED \cite{SSP:SGU:OR}. As an illustrative example
we consider one-component scalar theory with double-well potential. In this
example, we study some features of proposed approach, in particular, the
selection problem of stable vacuum state, what allows to avoid appearance of
tachyonic regimes. The similar analysis is possible for some other models of
such kind: the Friedberg-Lee model \cite{SSP:SGU:FL}, the non-linear 
$\eta$  meson model of Witten--Di Vecchia--Veneziano
\cite{SSP:SGU:Ahrensmeier, SSP:SGU:Blaschke}, end so on (e.g.,
\cite{SSP:SGU:Keiser,SSP:SGU:Hama}).

We use the metric $g^{\mu\nu} = diag(1,-1,-1,-1)$ and assume $\hbar = c =
1$.

\section{System of Basic Equations}

Let us consider the   scalar particle Lagrangian with  a self-interaction %
\begin{eqnarray}
\mathcal{L}[\Phi] & = &\frac{1}{2}\partial_{\mu}\Phi\partial^{\mu}\Phi-
\frac{1}{2}m_{0}^{2}\Phi^{2}-V[\Phi].\label{lag1}
\end{eqnarray}
It is assumed that the field variable $\Phi$ can be decomposed into the
classical space homogeneous background field $\phi_{0}(t)$ and  the fluctuation
part $\phi(x)$,
\begin{equation}
\Phi(x)=\phi_{0}(t)+\phi(x),\qquad
\langle\phi(x)\rangle =0,\qquad
\langle\Phi(x)\rangle=\phi_{0}(t),\label{back}
\end{equation}
where symbol $\langle\ldots\rangle$ denotes some averaging procedure.

The potential energy expansion in powers of $\phi(x)$ can be performed:
\begin{eqnarray}
V[\Phi]=V[\phi_{0}]+R_{1}[\phi_{0}]\phi+
\frac{1}{2}R_{2}[\phi_{0}]\phi^{2}+Q[\phi_{0},\phi],\label{dec}
\end{eqnarray}
where $R_{1}[\phi_{0}]$ and $R_{2}[\phi_{0}]$ are the first and the second
coefficients of the expansion, $Q[\phi_{0},\phi]$ contains high order terms
and will be neglected in the current article (non-dissipative approximation).
The equation of motion after field decomposition (\ref{back}) is (dots
denote derivatives with respect to time):
\begin{eqnarray}\label{mot}
[\,\partial_{\mu}\partial^{\mu}+m^{2}(t)\,]\phi+
\ddot{\phi}_{0}+m_{0}^{2}\phi_{0}+R_{1}[\phi_{0}]+
\frac{1}{2}\frac{dR_{2}[\phi_{0}]}{d\phi_{0}}\,\phi^{2}=0,
\end{eqnarray}
where $m(t)$ is the time dependent mass,
\begin{equation}\label{mass}
 m^{2}(t)=m_{0}^{2}+R_{2}[\phi_{0}].
\end{equation}
After averaging of the Eq.(\ref{mot}) we get the equation of motion for
the background field:
\begin{equation}
\ddot{\phi}_{0}+m_{0}^{2}\phi_{0}+R_{1}[\phi_{0}]+
\frac{1}{2}\frac{dR_{2}[\phi_{0}]}{d\phi_{0}}\langle\phi^{2}\rangle
=0.\label{mean}\end{equation}

The kinetic equation(KE) for the fluctuations can be obtained either by the
Bogoliubov transformation \cite{SSP:SGU:we,SSP:SGU:Grib} or by the
oscillator representation approach \cite{SSP:SGU:OR}. We found the last one
more efficient in application to the
 current model.

The derivation of KE is fulfilled in the quasiparticle representation, in
which the corresponding quadratic Hamiltonian has the diagonal form. One can
find the detailed formalism in the works \cite{SSP:SGU:we,SSP:SGU:OR}.The
resulting KE  has the following form:
\begin{eqnarray}
\frac{df(\bar{p},t)}{dt}=\frac{1}{2}\Delta(\bar{p},t) \int_{-\infty}^{t}dt'
\Delta(\bar{p},t')[1+2f(\bar{p},t)]
\cos\left[2\int_{t'}^{t}d\tau\omega(\bar{p},\tau)\right],\label{ke}
\end{eqnarray}
where $f(\bar{p},t)$ is the distribution function of particles and the
transition amplitude is given by
$\Delta(\bar{p},t)=\dot{\omega}(\bar{p},t)/\omega(\bar{p},t)$, where
$\omega(\bar{p},t)=\sqrt{m^2(t) +\bar{p}^2}$.

The KE (\ref{ke}) can be transformed to a system of ordinary differential
equations, which is convenient for numerical analysis\cite{SSP:SGU:we}
\begin{eqnarray}
2 \dot{f}=\Delta v_1,\qquad\dot{v}_1=\Delta (1+2f)-2\omega v_2,\qquad
\dot{v}_2=2\omega v_1 .\label{31}
\end{eqnarray}
Finally, the average value $\langle in|\phi^{2}(x)|in\rangle$ from Eq.
(\ref{mean}) is equal to
\begin{eqnarray}
\label{33} \langle in|\phi^2(x)|in\rangle = \int \frac{ d^3p }{2\omega(\bar
p, t)} \left[ 1+2f(\bar p, t) + v_1(\bar p, t) \right].
\end{eqnarray}
As a result, the equation of motion for background field Eq.(\ref{mean})
will have the form (the vacuum unit is omitted here)
\begin{equation}\label{cond}
\ddot{\phi}_{0}+m_{0}^{2}\phi_{0}+R_{1}[\phi_{0}]+
\frac{1}{2}\frac{dR_{2}[\phi_{0}]}{d\phi_{0}} \int \frac{ d^3p }{\omega(\bar
p, t)} \left[ f(\bar p, t) + \frac12 v_1(\bar p, t) \right] =0.
\end{equation}

The KE (\ref{ke}) (or equivalent equation system (\ref{31})) and
Eq.(\ref{cond}) form the closed system of equations of back
reaction problem. For a description of vacuum particle creation we use the
total particle density
\begin{equation}
n(t)=\int\frac{d\bar{p}}{(2\pi)^{3}}f(\bar{p},t),\label{dens}
\end{equation}
as well as condensate and created particle energies defined as
\begin{eqnarray}
E_{c} = \frac12 {\dot\phi}_0^2 + \frac12 m_0^2 \phi_0^2 +
\frac{1}{4} \lambda \phi_0^4, \quad E_{q} = \int\frac{d^3p}{(2\pi)^{3}}
\omega(\bar p, t) f(\bar{p},t).
\end{eqnarray}

\section{Examples: some polynomial potentials}

\noindent \textbf{3.1}. As the first example here we consider the $\Phi^4$
theory $V[\Phi]=\lambda\Phi^{4}/4,\ \lambda>0$. In this case we have
$R_1=\lambda\phi_0^3$, $R_2=3\lambda\phi_0^2$ and the time dependent mass
$m^2(t)=m_0^2+3\lambda\phi_0^2$. The Eq.(\ref{cond}) is
\begin{equation}\label{me}
\ddot\phi_0+\left(m_0^2+3\lambda\int\frac{d^3p}{\omega(t)}\left[f+ \frac12
v_1\right]\right)\phi_0+\lambda\phi_0^3=0.
\end{equation}

\begin{figure}[h]
\centering
\includegraphics[width=65mm,height=55mm]{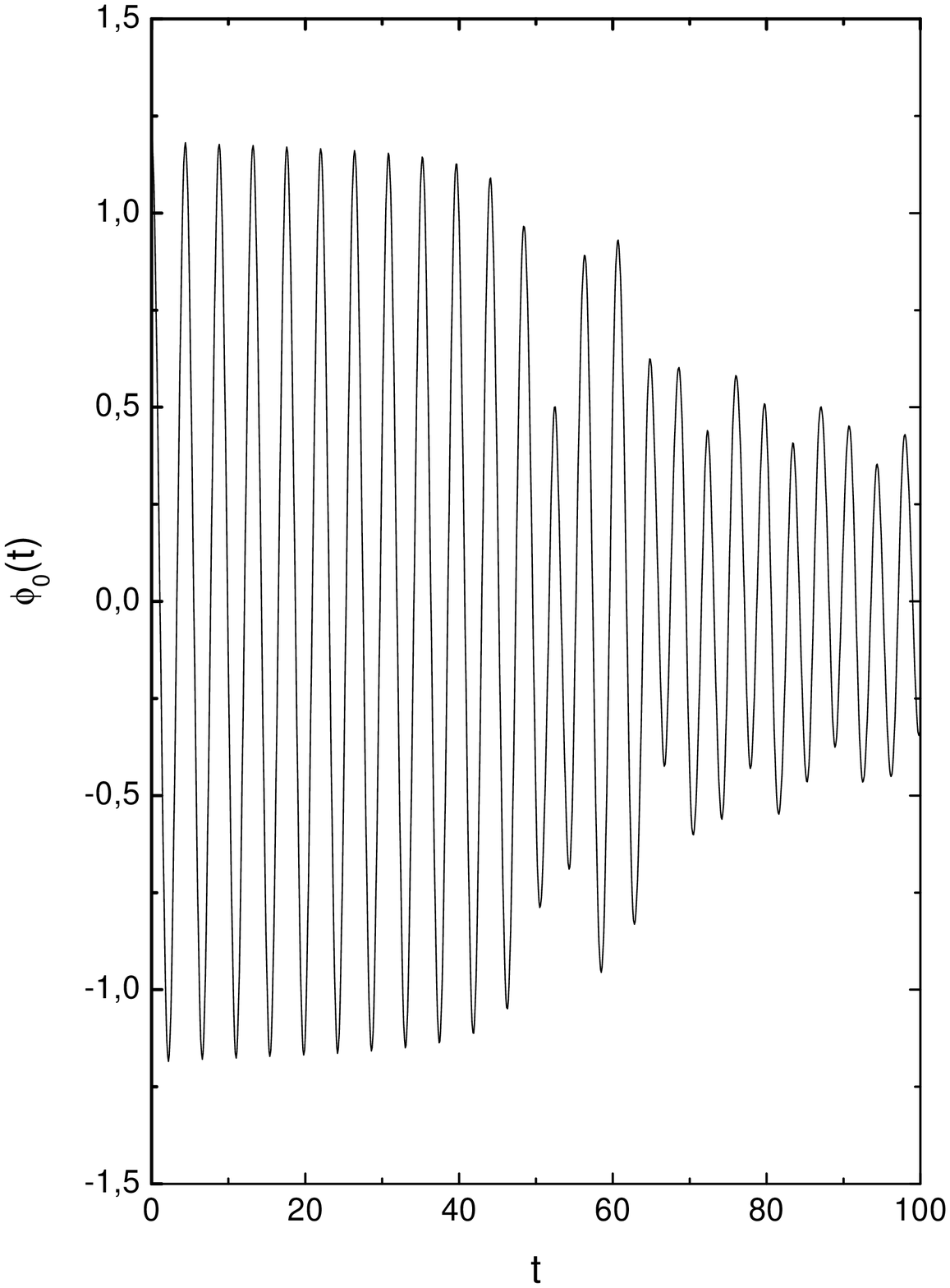}
\includegraphics[width=65mm,height=55mm]{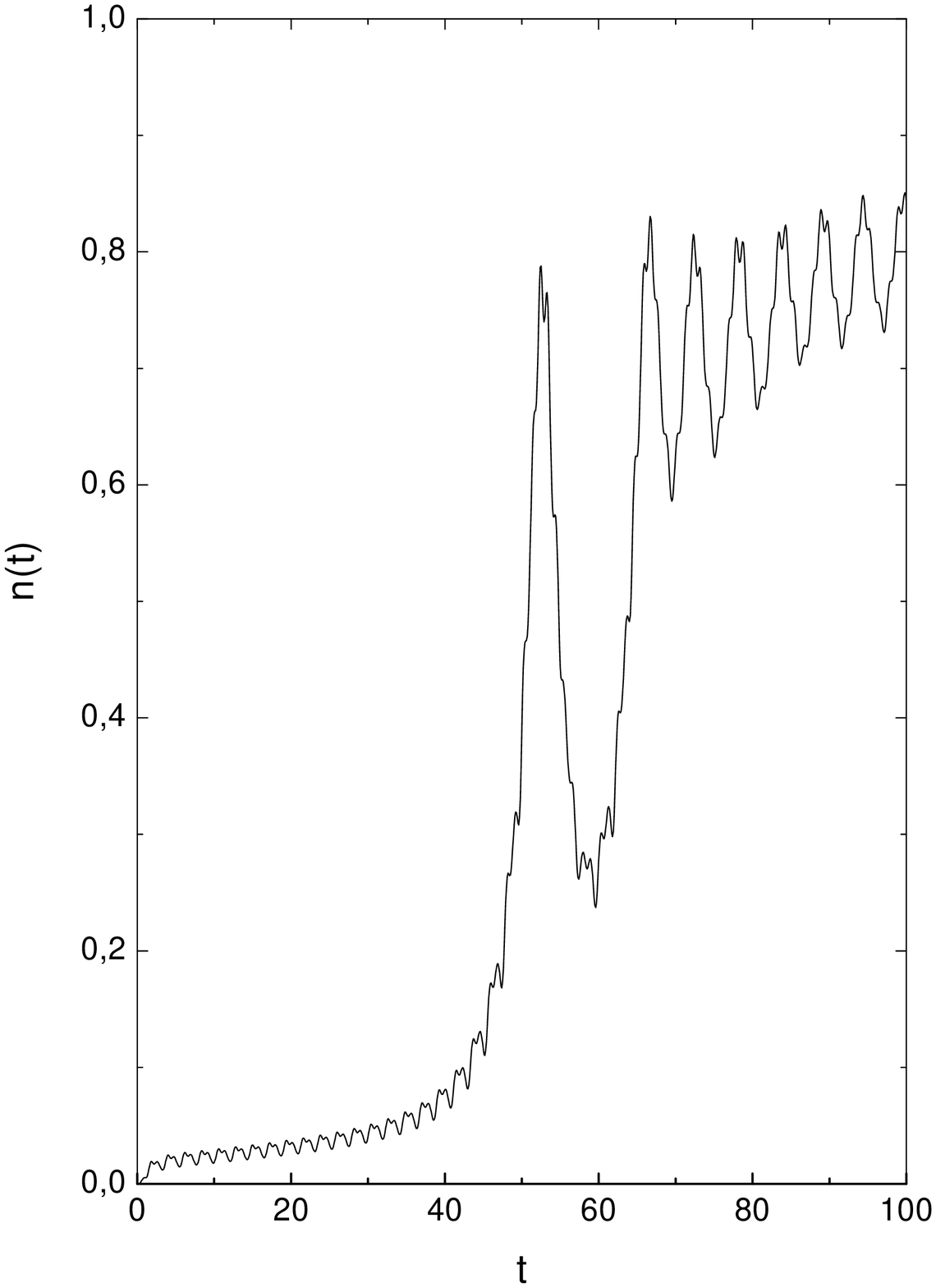}
\includegraphics[width=65mm,height=50mm]{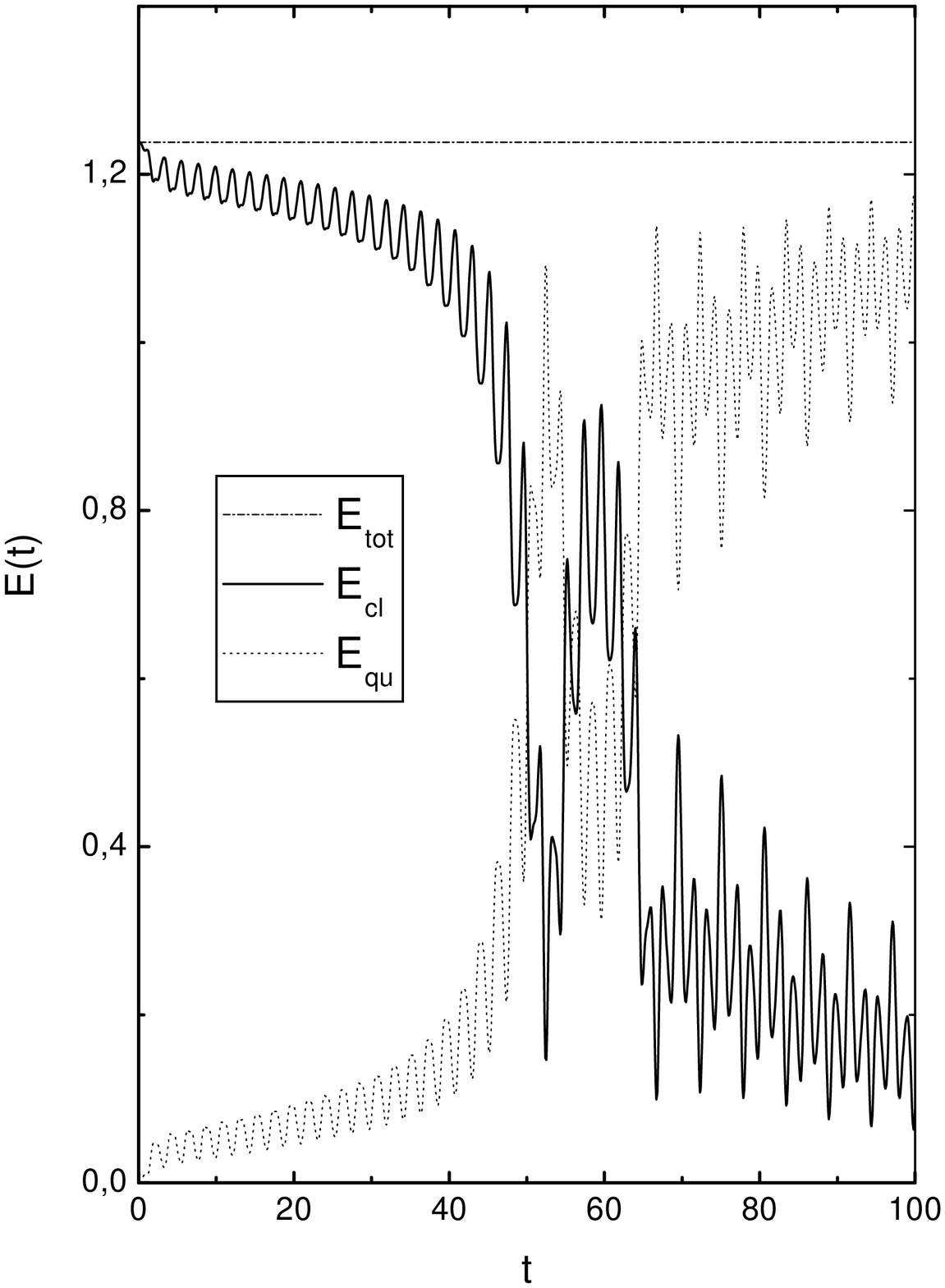}
\includegraphics[width=65mm,height=50mm]{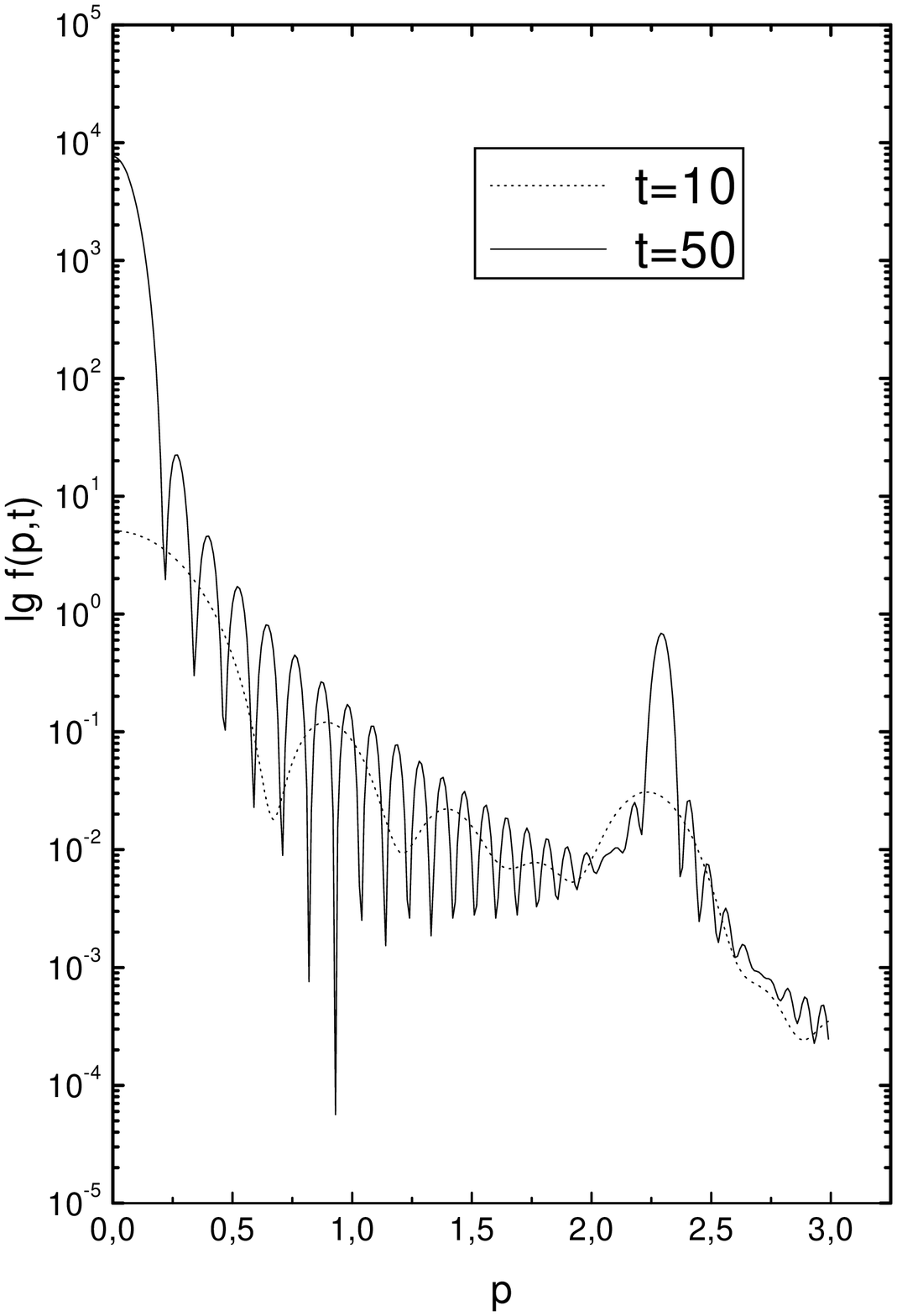}
\caption{Time evolution for the symmetric $\Phi^4$ potential. Parameters:
$m_0=1$, $\phi_0(0)=1.2$, $\lambda=1$; (a) evolution of the mean field; (b)
evolution of the particle density; (c) evolution of the energy; (d) momentum
spectrum of particles at time $t=10$ and $t=50$. }
\end{figure}

In numerical calculations we apply the
zero initial conditions for the distribution function and nonzero one for the
background field $\phi_{0}(t_{0})=1.2$. The choice of parameters
($\lambda=1$ and $m_0=1$) as well as initial conditions is motivated by the
desire to make a comparison between our work and \cite{SSP:SGU:Baacke},
where authors offered the alternative method for description of quantum
systems under strong background field action.

As it can be seen on Fig.1 at the early evolution stage all energy is mainly
concentrated in the field oscillation. For $t < 50$ we have a slow growing
of number density. However it drastically increases at $t \sim 50$ and after
this time the quantum energy dominates over classical one. Note that full
energy of the system is conserved precisely.
Fig.1 also shows that the spectra of distribution function is close
to quasi-equilibrium one. Thus fluctuation temperature can be calculated,
but it is beyond the scope of this paper.

\noindent \textbf{3.2}. The double-well potential
$\lambda\Phi^4/4-\mu^2\Phi^2/2$ results in the time dependent masses of
quasiparticles and background field excitations, which are defined by
Eq.(\ref{mass}) with substitution $m_0^2\to -\mu^2$. The stable
(non-tachyonic) solutions are absent in the neighbourhood of the point $\Phi=0$.
The transition of the system with broken symmetry to the stable state  is realized by the
shift at one of two stable points, $\Phi=\Psi_{\pm}+\Psi$, with the
following selection of the background field, $\Psi = \phi_0+\phi$, where
$\Psi_{\pm}=\pm \mu/\sqrt{\lambda}$. 
After these transformations we have:
\begin{equation}\label{me1}
  \ddot\phi+2\mu^2\phi_0+3\lambda\Psi_{\pm}\phi_0^2+\lambda\phi_0^3+
  3\lambda(\Psi_{\pm}+\phi_0)\int\frac{d^3p}{\omega(\bar p,t)}\left[f
  +\frac12 v_1\right]=0
\end{equation}
and the transition amplitude in the KE
\begin{equation}\label{}
  \Delta(\bar p,t)=\frac{3}{2}\lambda \dot\phi_0
  (\phi_0+\Psi_{\pm})\ \omega^{-2}_{\pm}(\bar p ,t),
\end{equation}
where $\omega_{\pm}$ are defined by the time dependent masses
\begin{equation}
m_{\pm}^2(t)=m_0^2+2\mu^2+3\lambda\phi_0(\phi_0+2\Psi_{\pm})\label{massw}.
\end{equation}
It is seen from Eqs.(\ref{me1}) and (\ref{massw}) that this redefinition of
vacuum state leads to the stable (non-tachyonic) evolution of the system.
Detailed research of these basic equations will be done in the future.

This work was partly supported by a Russian Federations State Committee for
Higher Education grant(E02-3.3-210) and RFBR grant(03-02-16877).

We are grateful to J. Baacke and D. Blaschke for interest to this work.


\begin{thebibliography}{99}\itemsep=0pt

\bibitem{SSP:SGU:Keiser} D. I. Kaiser, Phys.Rev.~D \textbf{59}, 117901 (1999).

\bibitem{SSP:SGU:Ahrensmeier}
D. Ahrensmeier, R. Baier, and D. Dirk, Phys. Lett.~ B \textbf{484}, 58
(2000).

\bibitem{SSP:SGU:Blaschke}
D. B. Blaschke, F. M. Saradzev, S. M. Schmidt, and D. V. Vinnik, Phys.
Rev.~D. \textbf{65}, 054039 (2002).

\bibitem{SSP:SGU:we}  S. A.~Smolyansky, G.~R\"opke, S. M.~Schmidt, D.~Blaschke,
V. D.~Toneev, and A. V.~Prozorkevich, hep-ph/9712377;
S. M. Schmidt, D.~Blaschke, G.~R\"opke, S. A.~Smolyansky,
A.~V.~Prozorkevich, and V. D.~Toneev, Int.~J.~Mod.~Phys.~E \textbf{7}, 709
(1998).

\bibitem{SSP:SGU:Baacke} J. Baacke and A. Heinen, Phys.Rev.~D \textbf{67}
  105020 (2003).

\bibitem {SSP:SGU:Grib} A. A. Grib, S. G. Mamaev, and V. M. Mostepanenko,
\textit{Vacuum Quantum Effects in Strong External Fields}, Friedmann
Laboratory Publishing, St. Peterburg, 1994.

\bibitem {SSP:SGU:Birrel} N. D. Birrell and P. C. W. Davies,
{\it Quantum Fields in the curved space}, Cambridge University Press, 1984.

\bibitem {SSP:SGU:Kluger} Y. Kluger, J. M. Eisenberg, B. Svetitsky, F.
Cooper, and E. Mottola, Phys. Rev. Lett. \textbf{67}, 2427 (1991); Phys.
Rev.~D \textbf{45}, 4659 (1992); \textbf{48},190 (1993).


\bibitem{SSP:SGU:OR} V. N. Pervushin, V. V. Skokov, A. V. Reichel,
S. A. Smolyansky, and A.~V.~Prozorkevich, hep-ph/0307200.

\bibitem{SSP:SGU:FL} R. Friedberg and T. D. Lee, Phys. Rev.~D \textbf{15},
1694 (1977); \textbf{16}, 1096 (1977); S. Loh, C. Greiner, U. Mosel, and M.
H. Toma, Nucl. Phys.~A \textbf{619}, 321 (1977).

\bibitem{SSP:SGU:Hama} T. Hamazaki, M. Sasaki, T. Tanaka, and
K. Yamamoto, Phys.Rev.~D \textbf{53}, 2045 (1996).


\end{thebibliography}
\end{document}